\documentclass[aps,floats]{revtex4}
\usepackage{amsmath,amssymb}
\usepackage{graphicx,epsfig}

\begin{document}
\bibliographystyle {plain}

\def\oppropto{\mathop{\propto}} 
\def\opsimeq{\mathop{\simeq}}
\def\opoverderline{\mathop{\overline}}
\def\operarrow{\mathop{\longrightarrow}}
\def\opsim{\mathop{\sim}}

\def\fig#1#2{\includegraphics[height=#1]{#2}}
\def\figx#1#2{\includegraphics[width=#1]{#2}}


\title{Random walk in a two-dimensional self-affine random potential : \\
properties of the anomalous diffusion phase at small external force } 


 \author{ C\'ecile Monthus and Thomas Garel }
  \affiliation{ Institut de Physique Th\'{e}orique, CNRS and CEA Saclay,
 91191 Gif-sur-Yvette, France}

\begin{abstract}
We study the dynamical response to an external force $F$ for a particle
performing a random walk in a two-dimensional quenched random potential of Hurst exponent $H=1/2$.  We present numerical results on the statistics of first-passage times that satisfy closed backward master equations. We find that there exists a zero-velocity phase in a finite region of the external force $0<F<F_c$, where the dynamics follows the anomalous diffusion law $ x(t) \sim \xi(F) \ t^{\mu(F)} $. The anomalous exponent $0<\mu(F)<1$ and the correlation length $\xi(F)$ vary continuously with $F$. In the limit of vanishing force $F \to 0$, we measure the following power-laws : the anomalous exponent vanishes as  $\mu(F) \propto F^a$ with $a \simeq 0.6$ (instead of $a=1$ in dimension $d=1$), and the correlation length diverges as $\xi(F) \propto F^{-\nu}$ with $\nu \simeq 1.29$ (instead of $\nu=2$ in dimension $d=1$). Our main conclusion is thus that the dynamics renormalizes onto an effective directed trap model, where the traps are characterized by a typical length $\xi(F)$ along the direction of the force, and by a typical barrier $1/\mu(F)$. The fact that these traps are 'smaller' in linear size and in depth than in dimension $d=1$, means that the particle uses the transverse direction to find lower barriers.

\end{abstract}

\maketitle

\section{ Introduction }

In quantum dynamics, the discovery of Anderson localization 
fifty years ago \cite{anderson} has shown that disorder can completely
suppress transport over large distances.
For classical random walks, the presence of quenched disorder can also induce 
completely new transport behaviors with respect to pure systems in some
regions of parameters (see the reviews \cite{HK,HBA,jpbreview} 
and references therein). In the present paper, 
we study the dynamical response to an
 external force $\vec F$ for a particle 
performing a random walk  in a two-dimensional quenched random potential $U(\vec r)$.
The dynamics of the position  $\vec r(t)$ of the particle
is governed by the Langevin equation
\begin{eqnarray}
\frac{d \vec r}{dt} =  \vec F - \vec \nabla U(\vec r(t)) + \vec \eta (t)
\label{Langevin}
\end{eqnarray}
where $\vec \eta(t) $ is a white noise 
\begin{eqnarray} 
< \eta_i(t) \eta_j(t ') > = 2 T \delta(t-t ') \delta_{i,j}
\end{eqnarray}
and where the quenched random potential
$U(\vec r)$ is self-affine with some Hurst exponent $H$
\begin{eqnarray}
\overline{ \left[ U(\vec r) -U(\vec r \ ') \right]^2 }
\opsimeq_{ \vert \vec r - \vec r \ ' \vert \to \infty}
 \vert \vec r - \vec r \ ' \vert^{2H}
\label{correU2d}
\end{eqnarray}

The random-force Sinai model that corresponds to 
dimension $d=1$ and Hurst exponent $H=1/2$,
has been much studied
by various exact methods
(see for instance the review \cite{jpbreview} and references therein)
with the following conclusions:

(i) for the case $\vec F=0$, the dynamics is logarithmically-slow 
$\vert x(t) -x(0) \vert \sim (\ln t)^{2}$ ;

(ii)  for the case $\vec F \ne 0$, there exists various dynamical phase
transitions as a function of $F \equiv \vert \vec F \vert $
 \cite{kesten,derrida_pom,feigelman_vin,jpb_annphys}.
The first important transition concerns the velocity $V$
defined as the ratio between the thermally averaged displacement $<x(t)>$
and the time $t$ in the limit $t \to +\infty$
\begin{eqnarray}
<x(t)>  && \simeq  V(F) t +...  \ \ {\rm for } \ \ F>F_c   \nonumber \\
<x(t)>  && \simeq  t^{\mu(F)} X +... \ \ {\rm for } \ \ F<F_c
\label{munonself}
\end{eqnarray}
For $F>F_c$, the velocity $V(F)$ is finite and self-averaging 
with $V(F \to F_c)=0$.
For  $F<F_c$, the velocity vanishes $V=0$, and the leading term of the displacement
is governed by a continuously varying exponent $0<\mu(F)<1$ 
with $\mu(F \to 0)=0$ and $\mu(F \to F_c) \to 1$.
In addition, the rescaled variable $X=<x(t)>/t^{\mu(F)}$
remains distributed over the samples.
In this anomalous diffusion phase, 
the width of the diffusion front in a given sample in also of order $t^{\mu(F)}$
\begin{eqnarray}
\left( <x^2(t)>-<x(t)>^2 \right)^{1/2}  \sim  t^{\mu(F)} \ \ \  {\rm for } \ \ F<F_c
\label{munonselfwidth}
\end{eqnarray}
More detailed statements about the diffusion fronts can be found
in the review \cite{jpbreview}.
From the point of view of strong disorder 
renormalization \cite{sinairg,sinaibiasdirectedtraprg,review}, the  
logarithmic behavior $ x(t) \sim  (\ln t)^2 $
for $F=0$ corresponds to an ``infinite disorder fixed point'',
whereas the anomalous diffusion phase $x(t) \sim t^{\mu(F)}$
 corresponds to a ``strong disorder fixed point'' 
in the limit of small external force $F \to 0$ where $\mu(F) \to 0$.

In higher dimension $d>1$, exact results are not available anymore,
but the logarithmic scaling at zero external force 
is still expected from scaling arguments on barriers \cite{Mar83,jpbreview}
\begin{eqnarray}
 \vert \vec r(t) - \vec r(0) \vert \sim (\ln t)^{1/H} \ \ {\rm for } \ \ F=0
\label{logslow}
\end{eqnarray}
We have recently checked numerically this behavior by various methods
in dimension $d=2$ \cite{sinai2d,conjugate,firstpassage}.
In the review \cite{jpbreview}, the response to an external force $F$
in dimension $d>1$ was left as an open question. We are not aware of
works on this question since. The aim of the present paper is thus to study
numerically the response to an external force in dimension $d=2$, and to determine in particular whether there exists a zero-velocity phase as in dimension $d=1$.
We will focus on observables that satisfy closed backwards equations and their
corresponding exact renormalization rules \cite{firstpassage}.
The fact that first-passage times (or similar observables like escape probabilities through a given boundary)
satisfy 'backwards master equation'
 is of course very well-known and can be found
in most textbooks on stochastic processes (see for instance 
\cite{gardiner,vankampen,risken,redner}). 
In the field of disordered systems, the backward Fokker-Planck equation
has been very much used to characterize
the dynamics of a single particle in a random medium (see for instance \cite{hernandez,Comtet_Dean,Dean_Maj,Maj_Comtet,Maj_Comtet2,Maj_Comtet3}),
but to the best of our knowledge, this approach has not much been used in higher dimension, nor for many-body problems. In our recent work \cite{firstpassage},
we have presented numerical results for a mean-field spin model, and 
the present two-dimensional random walk model without external force, i.e.
for the case $F=0$ in Eq. \ref{Langevin}. The present paper is devoted
to the case of a non-zero external force $F>0$.

The paper is organized as follows.
In section \ref{sec-master}, we define more precisely the model used in our numerical study. Section \ref{sec_escape} concerns escape probabilities,
whereas section \ref{sec_crossing} is devoted to first-passage times.
In section \ref{sec-trap}, we discuss the interpretation of our results
in terms of a renormalized directed trap model, and we discuss the 
similarities and differences with the same model in dimension $d=1$.
Our conclusions are summarized in section \ref{sec-conclusion}.

\section{ Master equation defining the stochastic dynamics }

\label{sec-master}

We consider a two-dimensional square lattice of size $L \times L$.
To generate a quenched random self-affine potential
 $U(\vec r)$ of Hurst exponent $H=1/2$ (Eq \ref{correU2d}),
we have used the so-called Weierstrass-Mandelbrot function method
(see \cite{sinai2d} and references therein for more details).
In the presence of an external force $F$ in the $x$-direction,
it is useful to introduce the tilted potential $U_F$ defined as
\begin{eqnarray}
U_F( x,y ) \equiv U(x,y) - F x
\label{UavecF}
\end{eqnarray}

We consider the following master equation that describes
 the time evolution of the
probability $P_t ({\vec r} ) $ to be at  position ${\vec r}$
 at time t 
\begin{eqnarray}
\frac{ dP_t \left({\vec r} \right) }{dt}
= \sum_{\vec r \ '} P_t \left({\vec r}\ ' \right) 
W \left({\vec r}\ ' \to  {\vec r}  \right) 
 -  P_t \left({\vec r} \right) W_{out} \left( {\vec r} \right)
\label{master}
\end{eqnarray}
where $ W \left({\vec r}\ ' \to  {\vec r}  \right) $ 
represents the transition rate per unit time from position 
${\vec r}\ '$ to ${\vec r}$, and 
\begin{eqnarray}
W_{out} \left( {\vec r} \right)  \equiv
 \sum_{ {\vec r} '} W \left({\vec r} \to  {\vec r}\ ' \right) 
\label{wcout}
\end{eqnarray}
represents the total exit rate out of position ${\vec r}$.
We have chosen  
the Metropolis dynamics at temperature $T=1$
defined by the transition rates 
\begin{eqnarray}
W \left( \vec r \to \vec  r \ '  \right)
= \delta_{<\vec r, \vec r\ ' >} 
\  {\rm min} \left(1, e^{-  (U_F(\vec r \ ' )-U_F(\vec r ))/T } \right)
\label{metropolis}
\end{eqnarray}
in terms of the tilted potential of Eq. \ref{UavecF}.
The first factor $\delta_{<\vec r, \vec r\ ' >}$
 means that the two positions
are neighbors on the two-dimensional lattice.
Various numerical results concerning the case $F=0$ can be found in our recent works \cite{sinai2d,conjugate,firstpassage}.
In the following, we consider only the case $F>0$.

\section{ Statistics of the escape probability against the external force}

\label{sec_escape}

\subsection{ Backward master Equation for the escape probabilities }

\begin{figure}[htbp]
 \includegraphics[height=6cm]{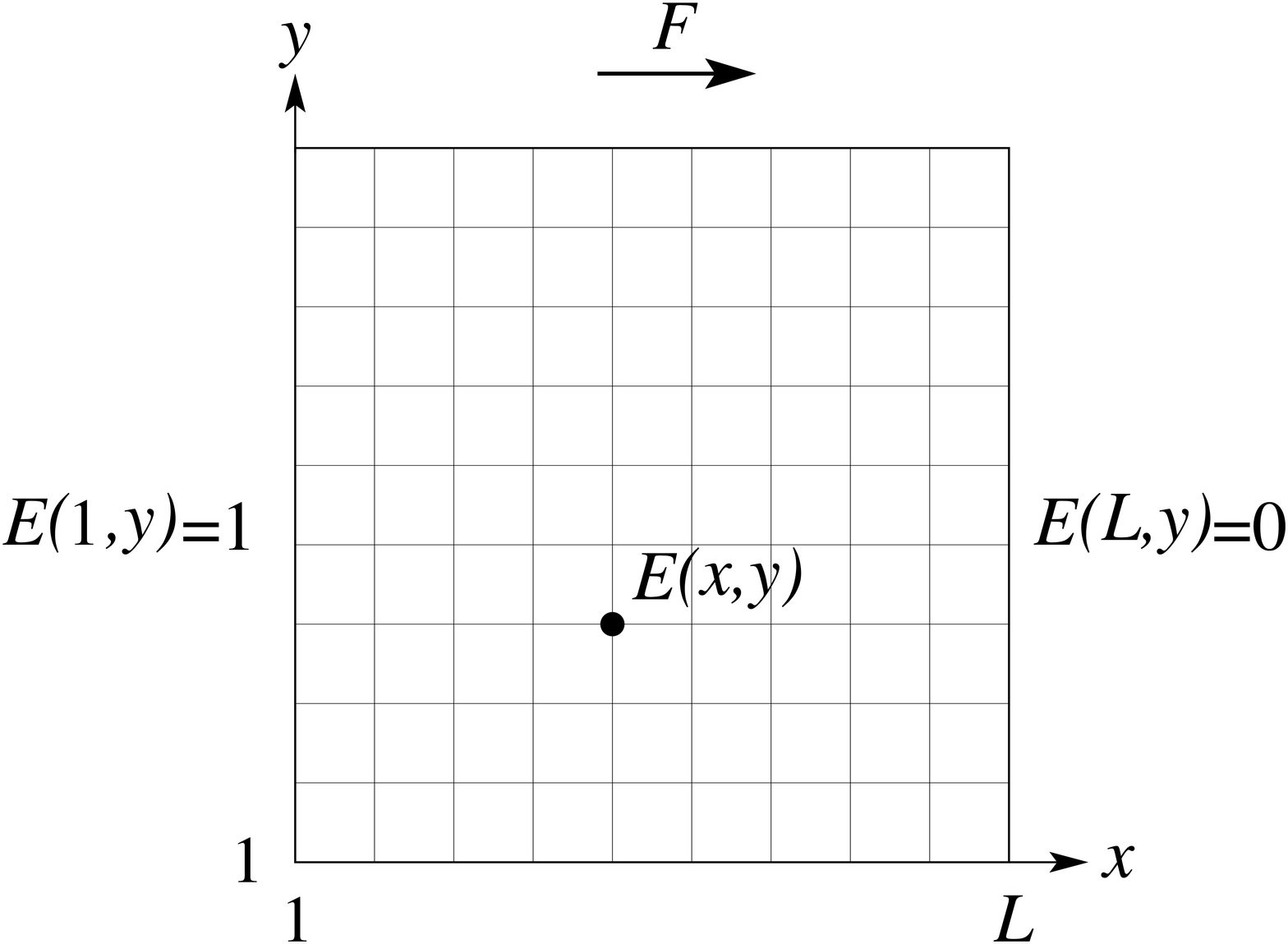}
\caption{ We consider the lattice continuous random walk in the random potential containing an external force $F$ in the $x$-direction (Eq. \ref{UavecF}). 
The escape probability $E(x,y)$, defined as the probability to reach first the boundary at $x=1$
before the boundary at $x=L$, when starting at position $(x,y)$,
 satisfies the backward master equation of Eq. \ref{backwardE} with the boundary conditions of Eq. \ref {bcE}. Below, we give numerical results on the 
typical behavior of 
$E_{center}=E(x=L/2,y=L/2)$ as a function of the size $L$ and of the external force $F$. }
\label{figescapedef}
\end{figure}

In this section, we consider the dynamics defined by the Master Equation of Eq. \ref{master}
on a square lattice of size $L \times L$, with absorbing boundaries
along the lines $x=1$ and along $x=L$, and reflecting boundaries
along the lines $y=1$ and along $y=L$.

The probability $E(\vec r)$ to reach first the absorbing boundary at $x=1$
(before the other absorbing boundary at $x=L$), 
when starting at position $\vec r$,
satisfies the 'backward master equation' \cite{gardiner,vankampen,risken}
\begin{eqnarray}
\sum_{\vec r \ '} 
W \left({\vec r}\  \to  {\vec r}\ '  \right) E \left({\vec r}\ ' \right) 
 -  E \left({\vec r} \right) W_{out} \left( {\vec r} \right) =0
\label{backwardE}
\end{eqnarray}
for $2 \leq x \leq L-1$ and $1 \leq y \leq L$, with 
the boundary conditions
\begin{eqnarray}
E(x=1,y) && =1 \ \ {\rm for } \ \ 1 \leq y \leq L  \nonumber \\
E(x=L,y) && =0 \ \ {\rm for } \ \ 1 \leq y \leq L
\label{bcE}
\end{eqnarray}
as illustrated on Fig. \ref{figescapedef}.

As explained in \cite{firstpassage}, 
if one eliminates iteratively the sites, the renormalized equations for the remaining sites keep the same form as Eq. \ref{backwardE} with renormalized rates
\begin{eqnarray}
\sum_{\vec r \ '} 
W^R \left({\vec r}\  \to  {\vec r}\ '  \right) E \left({\vec r}\ ' \right) 
 -   W_{out}^R \left( {\vec r} \right) E \left({\vec r} \right) =0
\label{rgbackwardE}
\end{eqnarray}
Upon the elimination of the site $\vec r_0$ via
\begin{eqnarray}
E \left({\vec r}_0 \right) =\frac{1}{W_{out}^R \left( {\vec r}_0 \right)}
 \sum_{\vec r \ ''} 
W^R \left({\vec r}_0\  \to  {\vec r}\ ''  \right) E \left({\vec r}\ '' \right) 
\label{elimEr0}
\end{eqnarray}
the renormalization rules read
\begin{eqnarray}
W^{R,new} \left({\vec r}\  \to  {\vec r}\ '  \right) && =  
W^R \left({\vec r}\  \to  {\vec r}\ '  \right)+ \frac{W^R \left({\vec r}\  \to  {\vec r}_0  \right) W^R \left({\vec r}_0\  \to  {\vec r}\ '  \right)}{W_{out}^R \left( {\vec r}_0 \right)} \nonumber \\
W_{out}^{R,new} \left( {\vec r} \right) && =W_{out}^R \left( {\vec r} \right)
- \frac{W^R \left({\vec r}\  \to  {\vec r}_0  \right) W^R \left({\vec r}_0\  \to  {\vec r}  \right)}{W_{out}^R \left( {\vec r}_0 \right)}
\label{rgrulesE}
\end{eqnarray}

In the following, we focus on the escape probability
\begin{eqnarray}
E_{center} \equiv E(\vec r_{center}) 
\label{defEcenter}
\end{eqnarray}
when starting at the center $\vec r_{center}=(x=L/2,y=L/2) $ of the square,
in order to characterize the asymmetry introduced by the external force $F$.
We may thus eliminate all sites in the region $2 \leq x \leq L-1$ 
except the single site $\vec r_{center}=(x=L/2,y=L/2)$.
Using Eq. \ref{bcE}, we finally obtain
\begin{eqnarray}
E_{center}=E(\vec r_{center}) = \frac{1}{W_{out}^R \left( {\vec r}_{center} \right)}
  \sum_{\vec r_S} W^R \left({\vec r}_{center}\  \to  {\vec r}_S  \right)
\label{Ecenter}
\end{eqnarray}
in terms of the set of points $\vec r_S=(x=1,1 \leq y \leq L)$
on the line $x=1$ where $E(\vec r_S)=1$

\subsection{ Numerical results on escape probabilities }

\begin{figure}[htbp]
 \includegraphics[height=6cm]{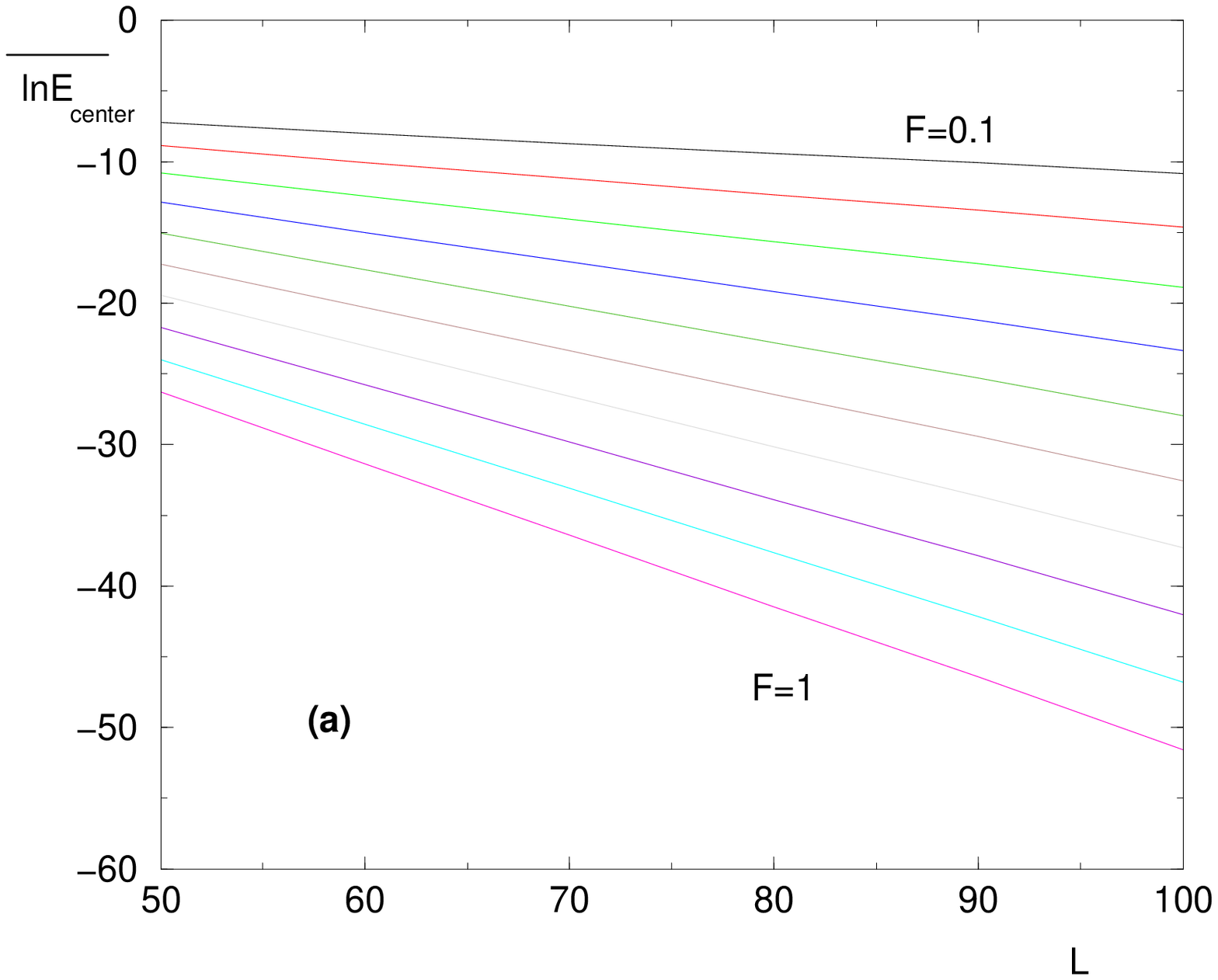}
\vspace{1cm}
 \includegraphics[height=6cm]{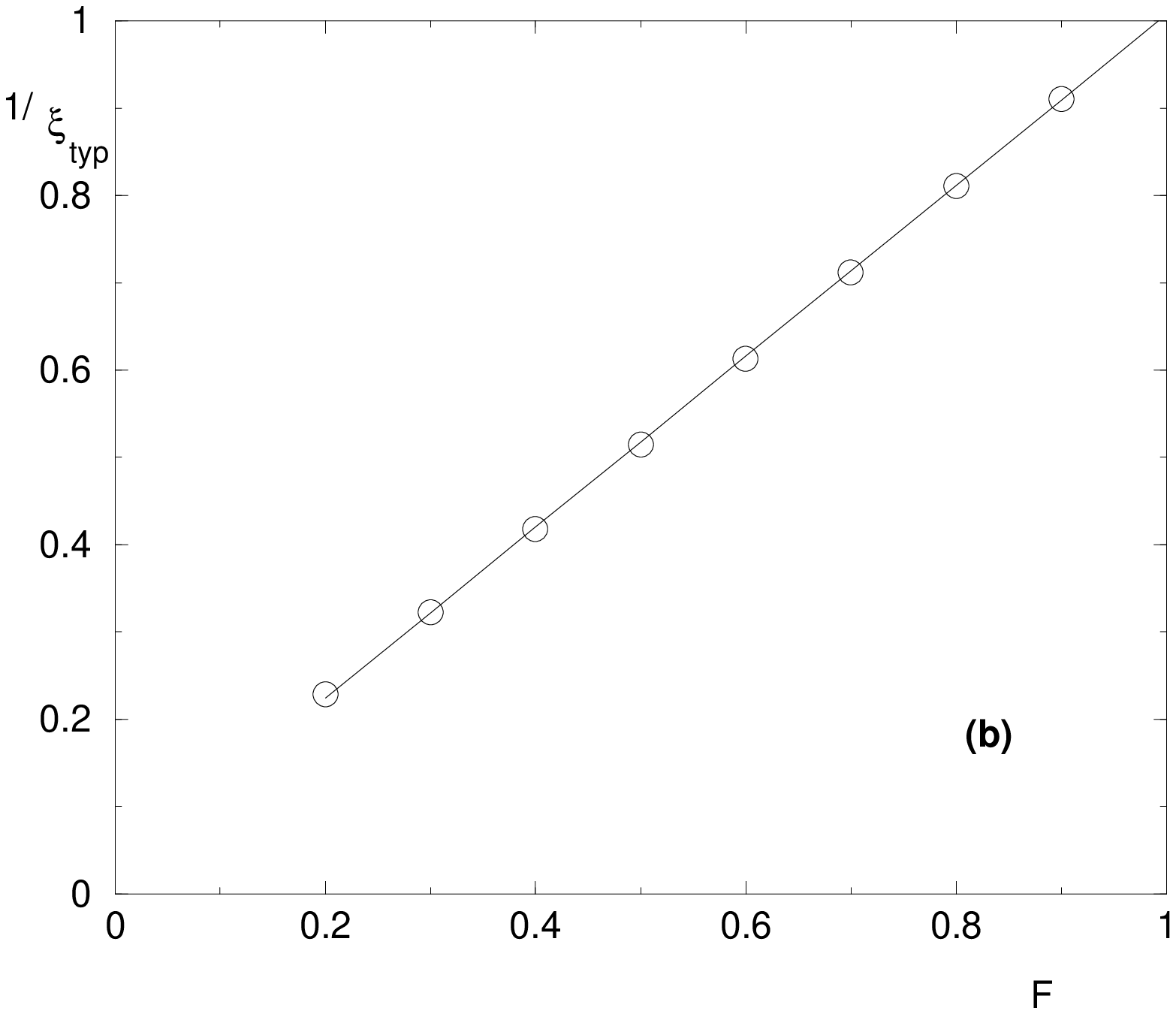}
\caption{ (Color on line) Typical behavior
of the escape probability $E_{center}$ defined in Eq. \ref{defEcenter} : 
(a) For each value of the force $0.1 \leq F \leq 1$, the typical value
$E_{center}^{typ}$ decays exponentially with the size $L$
(Eq. \ref{defxiF}).
(b) Inverse of the typical correlation length $\xi_{typ}(F)$ (Eq. \ref{defxiF})
as a function of the force $F$ : we obtain the linear behavior $\xi_{typ}(F)=1/F$. }
\label{figescape}
\end{figure}

For various values of the force $0.1 \leq F \leq 1$, we show on Fig. 
\ref{figescape} (a) the exponential decay with the size $L$
of the typical value
\begin{eqnarray}
E_{center}^{typ} \equiv e^{\overline{\ln E_{center}} }
\label{typEcenter}
\end{eqnarray}
This defines a typical correlation length $\xi_{typ}(F)$ via
\begin{eqnarray}
\ln E_{center}^{typ} \equiv \overline{ \ln E_{center}} 
\opsimeq_{L \to \infty} - \frac{(L/2)}{\xi_{typ}(F)} +...
\label{defxiF}
\end{eqnarray}
that will diverge in the limit of vanishing force
\begin{eqnarray}
\xi_{typ}(F) \oppropto_{F \to 0} \frac{1}{F^{\nu_{typ}}}
\label{defnutyp}
\end{eqnarray}
As shown on Fig. \ref{figescape} (b), the typical exponent is 
\begin{eqnarray}
\nu_{typ} =1
\label{nutyp1}
\end{eqnarray}
We find actually that even far from the $F \to 0$ limit, the typical 
correlation length is exactly given by
\begin{eqnarray}
\xi_{typ}(F) = \frac{1}{F}
\label{xityp}
\end{eqnarray}
This means that the correlation length seen in the escape probability $E_{center}$
directly reflects the linear asymmetry introduced by the force in
the tilted random potential of Eq. \ref{UavecF}.
So this length scale has exactly the same expression as in the one-dimensional case
\cite{sinairg,review} and in equivalent quantum spin chains \cite{dsf,review}.

However, it is well known that in one dimension there exists two different 
correlations lengths that diverge with different exponents as $F \to 0$ 
\cite{dsf,review}. In the present section, we have found that the typical correlation length is governed by the same exponent $\nu_{typ}=1$ as in $d=1$.
In the following section, we will study 
 the other correlation length $\xi(F)$ that appears in the statistics
of first-passage times.

\section{ Statistics of first-passage times }

\label{sec_crossing}

\subsection{ Backward master equation for the first-passage times }

\begin{figure}[htbp]
 \includegraphics[height=6cm]{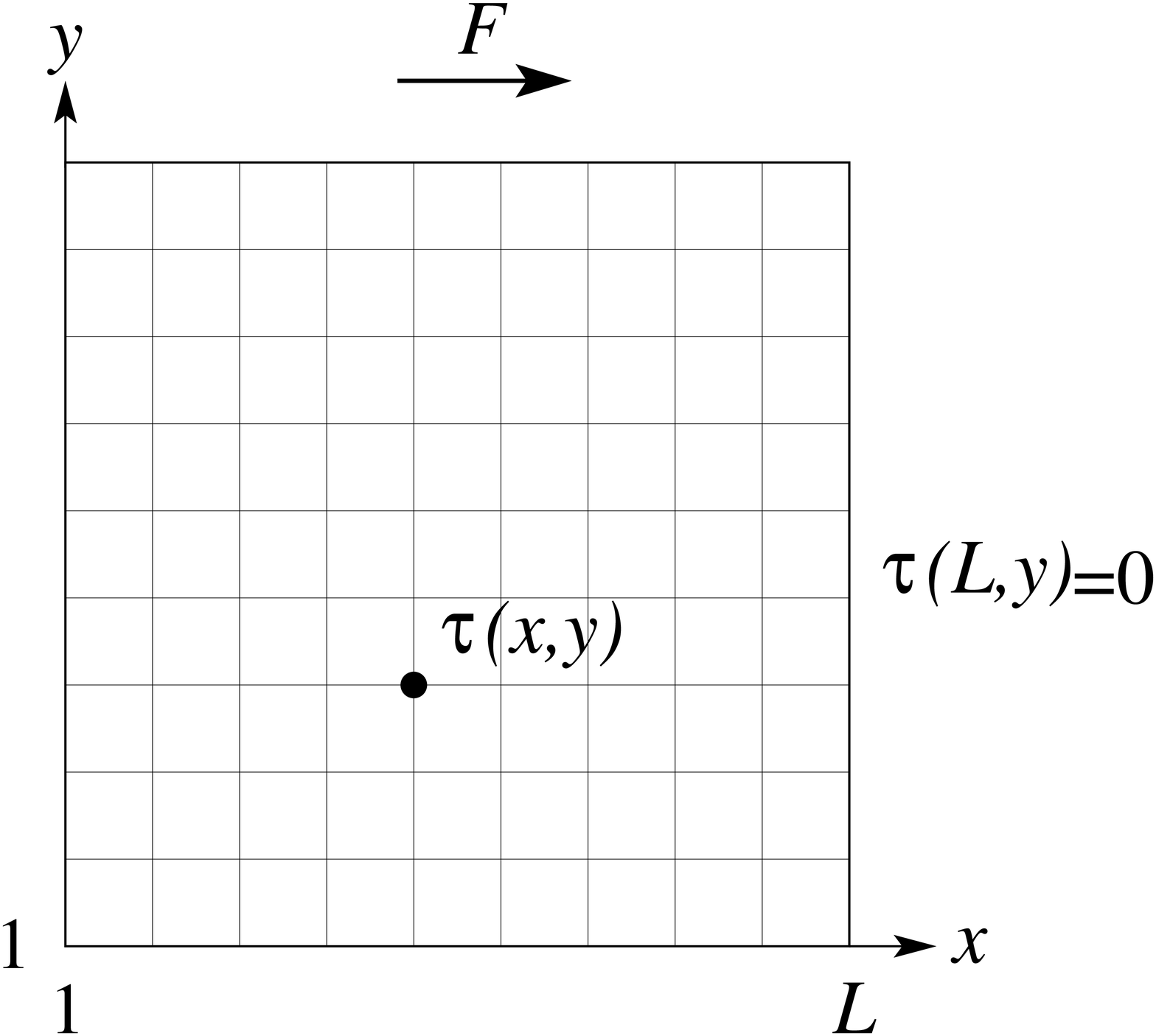}
\caption{ The mean first-passage time $\tau(x,y)$ on the boundary $x=L$
when starting at position $(x,y)$,
 satisfies the backward master equation of Eq. \ref{backwardtau} with the boundary conditions of Eq. \ref {bcT}.  Below, we give numerical results on the 
typical behavior of the crossing time 
$\tau_{cross}=\tau(x=1,y=L/2)$ as a function of the size $L$ and of the external force $F$. }
\label{figtimedef}
\end{figure}

Suppose the dynamics starts at time $t=0$ in position ${\vec r}$,
 and one is interested in the 
random time $t$ where the dynamics will 
reach for the first time any point on the line $x=L$.
As is well known 
(see for instance the textbooks \cite{gardiner,vankampen,risken,redner}),
the mean first-passage time $\tau(\vec r)=<t>$ 
(where the notation $<.>$ represents the average
 with respect to the dynamical trajectories) satisfies the following
'backward master equation' for all points outside the line $x=L$
\begin{eqnarray}
\sum_{\vec r \ '} 
W \left({\vec r}\  \to  {\vec r}\ '  \right) \tau \left({\vec r}\ ' \right) 
 -  W_{out} \left( {\vec r} \right)  \tau \left({\vec r} \right) = -1
\label{backwardtau}
\end{eqnarray}
whereas points all on the line $x=L$ satisfy the
 boundary condition 
\begin{eqnarray}
\tau(x=L,y )=0 \ \ {\rm for } \ \ 1 \leq y \leq L
\label{bcT}
\end{eqnarray}
as illustrated on Fig. \ref{figtimedef}.

Again as explained in \cite{firstpassage}, 
if one wishes to eliminate iteratively the sites, the 
renormalized equation will be of the form similar to Eq. \ref{backwardtau}
but with coefficients different from $(-1)$ on the right hand-side
\begin{eqnarray}
\sum_{\vec r \ '} 
W^R \left({\vec r}\  \to  {\vec r}\ '  \right) \tau \left({\vec r}\ ' \right) 
 -  W_{out}^R \left( {\vec r} \right) \tau \left({\vec r} \right) = -K^R\left({\vec r} \right)
\label{rgbackwardT}
\end{eqnarray}

Upon the elimination of the site $\vec r_0$
\begin{eqnarray}
\tau \left({\vec r}_0 \right) =\frac{1}{W_{out}^R \left( {\vec r}_0 \right)}
\left[ \sum_{\vec r \ ''} 
W^R \left({\vec r}_0\  \to  {\vec r}\ ''  \right) \tau \left({\vec r}\ '' \right) 
+ K^R\left({\vec r}_0 \right) \right]
\label{elimTr0}
\end{eqnarray}
the renormalization rules for the rates $W^R$ are the same as in Eq. \ref{rgrulesE}, whereas the new renormalization rule for
the inhomogeneous term of the backward master equation reads
\begin{eqnarray}
K^{R,new}\left({\vec r}\right)  = K^R\left({\vec r} \right)
+ \frac{W^R \left({\vec r}\  \to  {\vec r}_0  \right)}{W_{out}^R \left( {\vec r}_0 \right)} K^R\left({\vec r}_0 \right)
 \label{rgrulesT}
\end{eqnarray}

Here we wish to focus on the crossing time 
\begin{eqnarray}
\tau_{cross}=\tau(x=1,y=L/2) 
\label{defTcross}
\end{eqnarray}
representing the mean averaged time to reach any point of
the line $x=L$ when starting 
in the middle ($y=L/2)$ of the opposite line $x=1$.
Numerically, one can eliminate all points $x \leq L-1$ except the point
${\vec r}_{last}=(x=1,y=L/2)$. The crossing time is then computed as
\begin{eqnarray}
\tau_{cross}=\tau(x=1,y=L/2) =  \frac{K^R\left(x=1,y=L/2 \right)}{W_{out}^R \left( x=1,y=L/2 \right)}
\label{Tcross}
\end{eqnarray}

\subsection{ Numerical results on first-passage times}

\begin{figure}[htbp]
 \includegraphics[height=6cm]{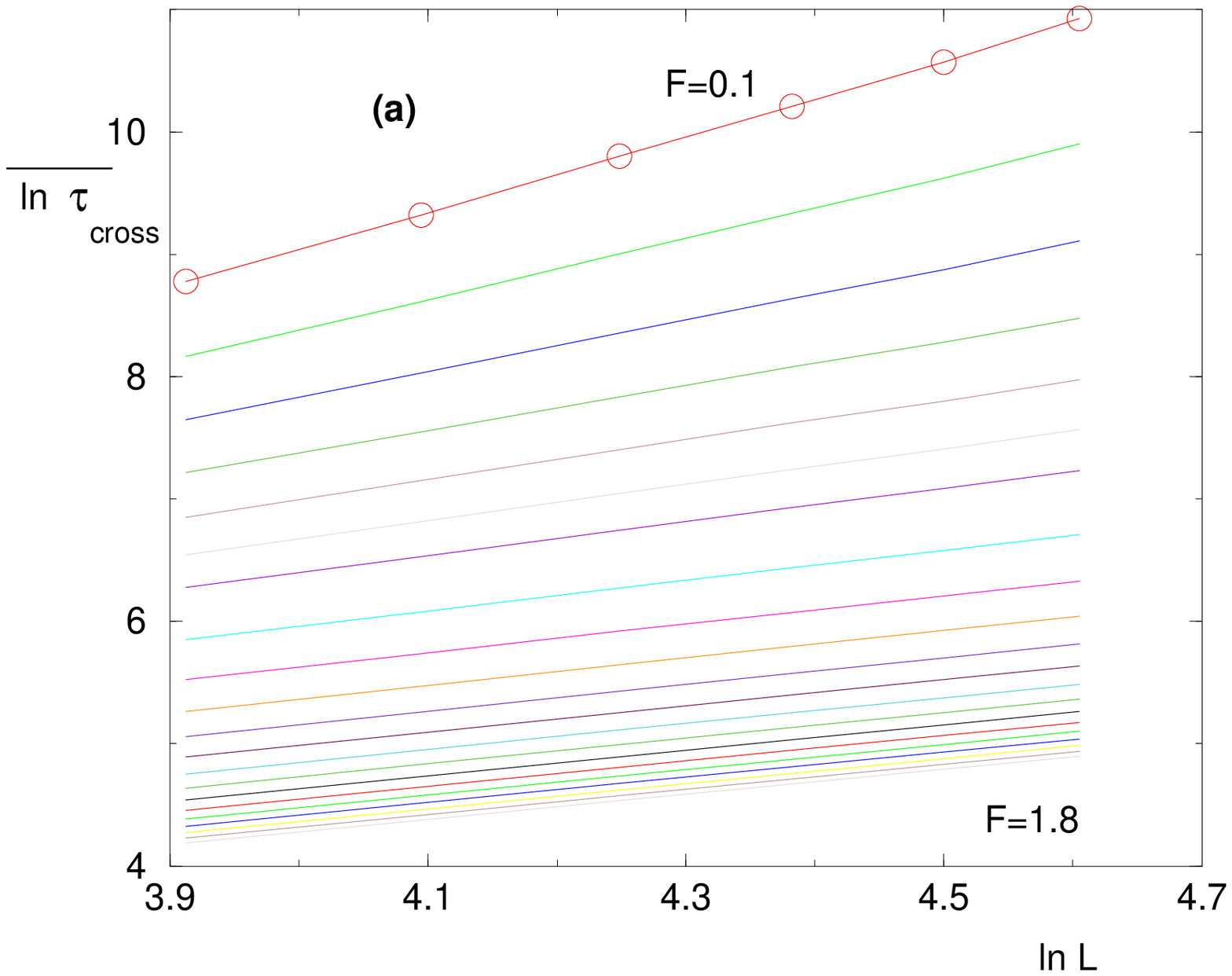}
\vspace{1cm}
 \includegraphics[height=6cm]{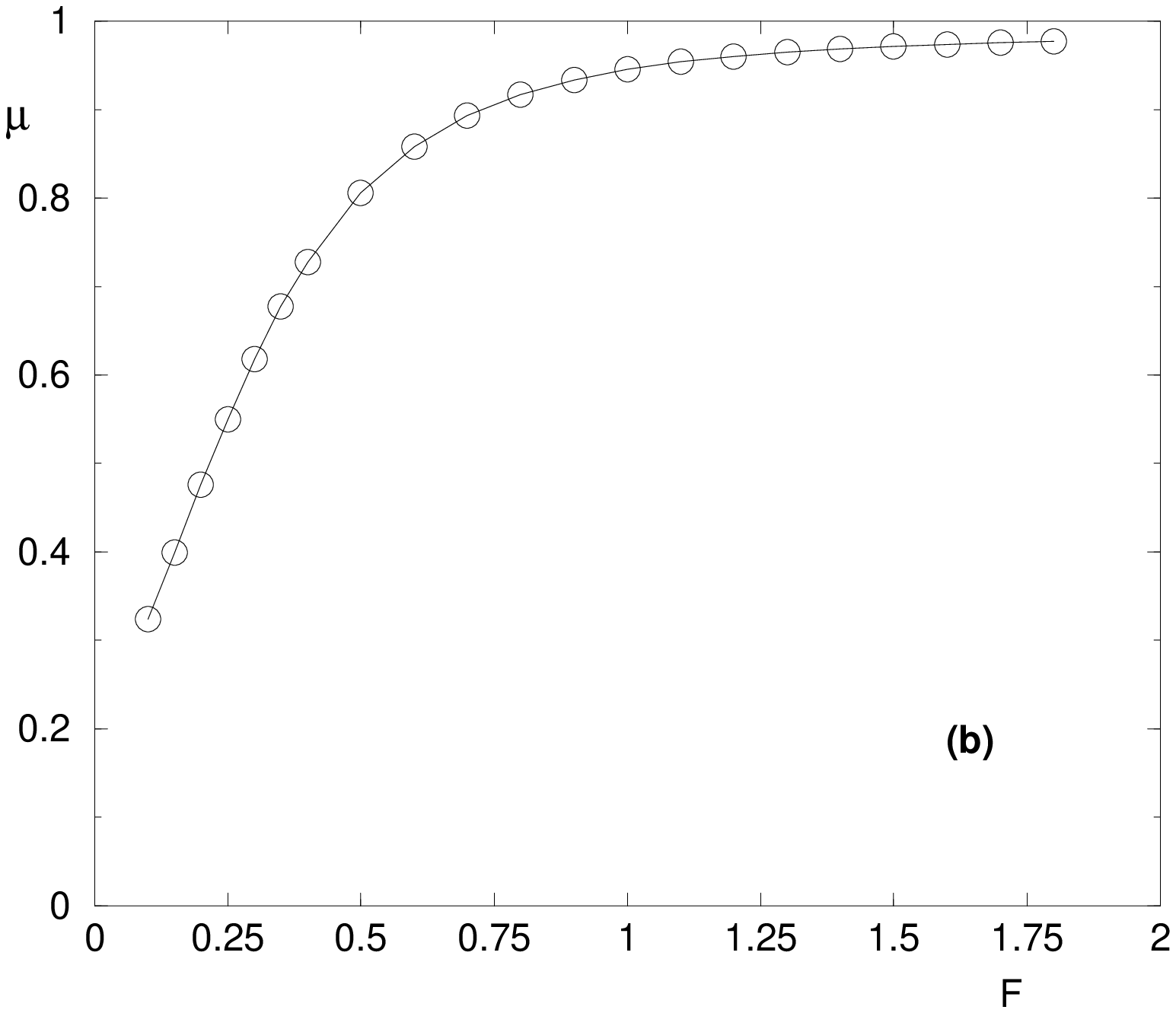}
\caption{(Color on line) Typical behavior
of the crossing time $\tau_{cross}$ defined in Eq. \ref{defTcross} :  
(a) For each value of the force $0.1 \leq F \leq 1.8$, 
$\tau_{cross}^{typ}$ grows as a power-law with the size $L$,
as shown here in log-log scale (see Eq. \ref{scalingTcross}).
(b) Corresponding anomalous exponent $\mu(F)$ defined in Eq. \ref{scalingTcross}
as a function of the force $F$.  }
\label{figfpt}
\end{figure}

\begin{figure}[htbp]
 \includegraphics[height=6cm]{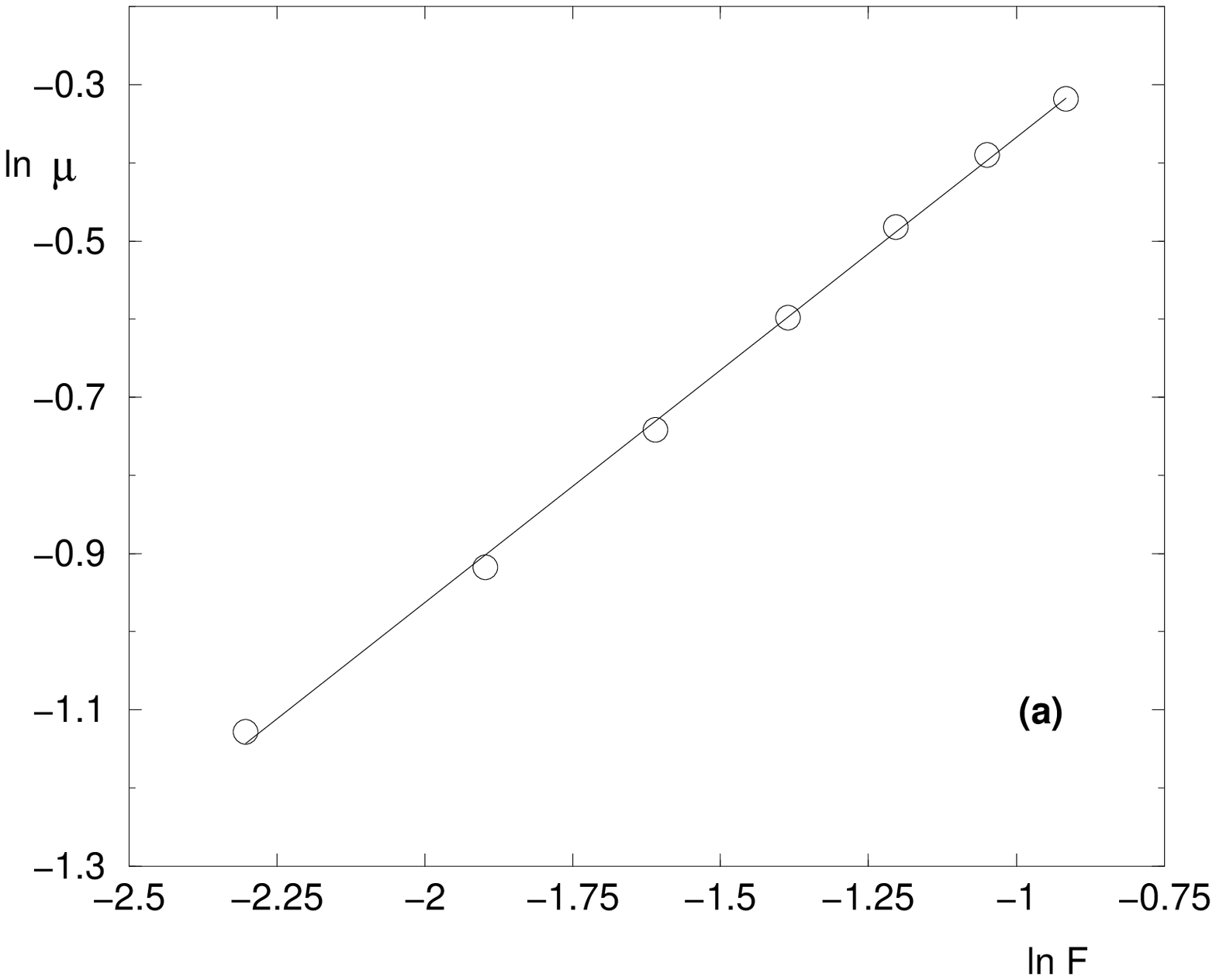}
\vspace{1cm}
 \includegraphics[height=6cm]{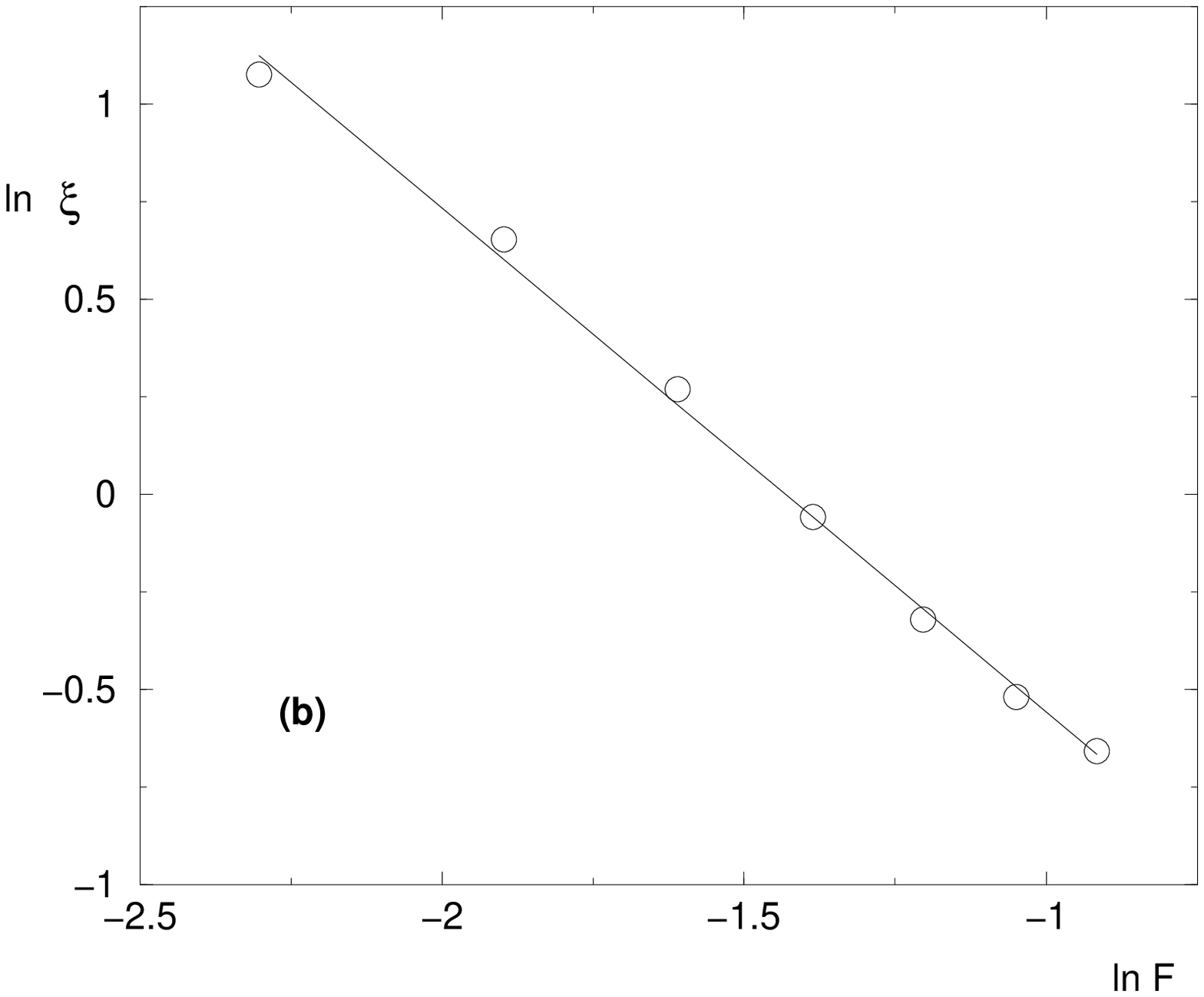}
\caption{ Critical exponents in the limit of vanishing force $F \to 0$
(the points correspond to $F=0.1,0.15,0.2,0.25,0.3,0.35,0.4$)
(a) Log-log plot of the exponent $\mu(F)$ as a function of $F$ 
 : the slope corresponds to $a \sim 0.6$ (see Eq. \ref{mucriti})
(b)  Log-log plot of the correlation length $\xi(F)$ as a function of $F$ 
 : the slope corresponds to $\nu \sim -1.29$  (see Eq. \ref{xicriti})}
\label{figexponents}
\end{figure}

As shown Fig. \ref{figfpt} (a), we find the following typical behavior in $L$
\begin{eqnarray}
\overline{ \ln \tau_{crossing}(F,L) }
 \simeq \frac{1}{\mu(F)} \left(\ln L - \ln \xi(F) \right)
\label{scalingTcross}
\end{eqnarray}
where the corresponding exponent $\mu(F)$ is shown as a function of
the external force $F$ on Fig. \ref{figfpt} (b).

After inversion, the behavior of the typical 
crossing time $\tau_{crossing}(F,L) $
 corresponds to the following dynamical behavior for the 
typical displacement $x(t)$
in the direction $x$ of the force, as a function of time $t$
\begin{eqnarray}
x_{typ}(t)  \sim \xi(F) \ t^{\mu(F)} 
\label{scalingxt}
\end{eqnarray}

Our results are thus compatible with a zero-velocity phase characterized
by an exponent $0<\mu(F)<1$ in a finite region $0<F<F_c$, 
and with a finite velocity phase
with $\mu(F)=1$ above the threshold $F>F_c$. 
We cannot give a very precise value for the threshold $F_c$
as a consequence of the rounding coming from finite-size effects.
Anyway, this critical value is not universal and depends on the details of
the random potential at short distances.
 As a consequence in the following, we will focus instead on the universal
properties that are expected to appear at small external force, where the dynamics is governed by the large-scale properties of the random potential.

In the region $0.1 \leq F \leq 0.4$, our results are compatible 
with the power-law
\begin{eqnarray}
\mu (F) \oppropto_{F \to 0} F^{a} \ \ {\rm with } \ \ a \simeq 0.6 
\label{mucriti}
\end{eqnarray}
as shown on Fig. \ref{figexponents} (a).
It is thus significantly different
from the exactly known value $a(d=1)=1$ for the same problem in dimension $d=1$ 
(see the review \cite{jpbreview} and references therein).
Similarly, as shown on Fig. \ref{figexponents} (b),
we find that the correlation length $\xi(F)$ 
appearing in the prefactor of Eq. \ref{scalingxt},
(that we measure from Eq. \ref{scalingTcross} by fitting our data shown on Fig.
\ref{figfpt} (a) diverges with the following power-law as $F \to 0$ 
\begin{eqnarray}
\xi (F) \oppropto_{F \to 0} F^{-\nu} \ \ {\rm with } \ \ \nu \simeq 1.29 
\label{xicriti}
\end{eqnarray}
Again this exponent is thus significantly smaller than the exactly known value $\nu(d=1)=2$ for the same problem in dimension $d=1$ (see the review \cite{jpbreview} and references therein).

\section{ Discussion : Equivalence with an appropriate renormalized directed trap model }

\label{sec-trap}

\subsection{ Scaling analysis of the equivalent problem in dimension $d=1$ }

As recalled in the introduction, the equivalent problem in dimension $d=1$
is exactly soluble by various methods. 
However, since these exact methods cannot be applied in dimension $d=2$, it is
important to identify the physical mechanism of the anomalous diffusion in
dimension $d=1$ to see if scaling arguments
 can be used in higher dimension $d>1$.
In dimension $d=1$, the physical picture is that the dynamics renormalizes
onto a directed trap model 
 \cite{jpbreview,sinairg,sinaibiasdirectedtraprg} 
with the following properties :

(i) the distribution $P(\tau)$ of trapping times decays as the power-law
\begin{eqnarray}
P(\tau) \oppropto_{\tau \to +\infty} \frac{1}{\tau^{1+\mu(F)}}
\label{ptau}
\end{eqnarray}
where the exponent $\mu(F)$ is exactly the exponent governing the anomalous diffusion law of Eq. \ref{scalingxt}. Equivalently, the distribution $P(B)$
of barriers $B \equiv \ln \tau$ against the drift $F$ decays exponentially
\begin{eqnarray}
{\cal P}(B) \oppropto_{B \to +\infty} e^{- \mu(F) B} 
\label{pB}
\end{eqnarray}
This means that $\mu(F)$ should be interpreted here as the inverse of the 
typical barrier 
\begin{eqnarray}
B_{typ}(F) = \frac{1}{ \mu(F) } 
\label{mubtyp}
\end{eqnarray}

(ii) the size $\xi(F)$ of a renormalized trap scales as 
\begin{eqnarray}
\xi(F) \propto \left( B_{typ}(F) \right)^{1/H} = \frac{1}{\left( \mu(F)\right)^{1/H} } 
\label{relationH}
\end{eqnarray}
as a consequence of the scaling property of the random potential
in terms of the Hurst exponent $H$ that yields $B_{typ}(F)= \left(\xi(F)\right)^H$. So the two exponents governing the divergences of barriers
\begin{eqnarray}
 B_{typ}(F) = \frac{1}{ \mu(F) } \oppropto_{F \to 0} \frac{1}{F^{a}} 
\label{defaarg}
\end{eqnarray}
and lengths
\begin{eqnarray}
\xi(F)  \oppropto_{F \to 0}\frac{1}{F^{\nu}} 
\label{defnuarg}
\end{eqnarray}
are related by the relation
\begin{eqnarray}
 \nu= \frac{a}{H}
\label{relationHexp}
\end{eqnarray}

(iii) the exponents $a$ and $\nu$ governing
the divergences of the typical barrier and of the typical size
in the limit of small external drift $F$ can be understood via
the following simple scaling argument in dimension $d=1$ :
the barrier $B(l)$ against the drift $F$ on a interval $l$ can be written as
\begin{eqnarray}
B(l) = -F l + l^H u
\label{resnuscaling}
\end{eqnarray}
where $u$ is a random variable of order $O(1)$.
The minimization of $B(l)$ with respect to $l$ yields the characteristic 
length scale 
\begin{eqnarray}
\xi^{(1d)}(F) \propto \frac{1}{F^{\nu_{1d}}} \ \ {\rm with } 
\ \ \nu_{1d}=\frac{1}{1-H}
\label{letoile}
\end{eqnarray}
above which the deterministic term $(-F l)$ becomes larger
than the random term in Eq. 
\ref{resnuscaling}. So $\xi^{(1d)}(F)$ represents the length of an effective renormalized trap, and the corresponding barrier scales as
\begin{eqnarray}
B_{typ}^{(1d)}(F) = B\left( l=\xi^{(1d)}(F)\right) \propto  \left( \xi^{(1d)}(F) \right)^H
\propto \frac{1}{F^{a_{1d}}}  \ \ {\rm with } 
\ \ a_{1d}=\frac{H}{1-H} = H \nu_{1d}
\label{Betoile}
\end{eqnarray}

In dimension $d=1$, the precise equivalence between the biased Sinai model
and the one-dimensional directed trap model is discussed in more details in
\cite{sinaibiasdirectedtraprg}. In particular, besides the scaling of 
Eq. \ref{scalingxt}, one can relate exactly the corresponding 
diffusion fronts that involve
 L\'evy stable distributions. We refer to \cite{sinaibiasdirectedtraprg} and references therein for more details.

\subsection{ Similarities and differences in dimension $d=2$ }

In dimension $d=2$, the numerical results presented above 
suggest that, as in dimension $d=1$, there exists an anomalous diffusion phase
as small external force, which is equivalent with a directed trap model
at large scales. The important difference is then in the properties
of the elementary effective renormalized traps, in particular in their
dependence upon the external force.
As explained in (i) and (ii) above, 
$\mu(F)$ should be interpreted as the inverse of the typical barrier
of an elementary trap, whereas $\xi(F)$ should be interpreted as its 
longitudinal length in the direction of the force.
Taking into account the numerical uncertainties on the 
the exponents $a$ and $\nu$ measured in section \ref{sec_crossing},
the values given in Eqs \ref{mucriti} and \ref{xicriti}
seem still 
compatible with the scaling relation of Eq. \ref{relationH} with $H=1/2$. 
Our conclusion is thus that
only the point (iii) which contains an explicit one-dimensional optimization 
is not valid anymore : we are not aware of the appropriate argument based
on a two-dimensional optimization that would give values of $a$ and $\nu$ in $d=2$.  The fact that these traps are 'smaller' as a function of $F$
than in the equivalent problem in dimension $d=1$ means that the particle uses the transverse direction
to find lower barriers. However, a quantitative analysis 
 of this effect goes beyond the present work.

\section{ Conclusion}

\label{sec-conclusion}

In this paper, we have studied the continuous-time random walk of a particle in a two-dimensional self-affine random potential of Hurst exponent $H=1/2$ in the presence of an external force $F$. We have presented numerical results on the statistics of escape probabilities and of first-passage times that satisfy closed backward master equations. Our main conclusion is that, as in dimension $d=1$,
 there exists a zero-velocity phase in a finite region of the external force $0<F<F_c$, where the dynamics follows an anomalous diffusion law $ x(t) \simeq \xi(F) \ t^{\mu(F)} $ with $0<\mu(F)<1$. However, 
in the limit of small external force $F \to 0$,
the exponents governing the vanishing
of the anomalous exponent $\mu(F) \propto F^a$ with $a \simeq 0.6$ 
and the divergence of
the correlation length $\xi(F)\propto F^{-\nu} $ with $\nu \simeq 1.29$
are significantly different from their exactly known values $a(d=1)=1$
and $\nu(d=1)=2$ in dimension $d=1$.
Finally, we have proposed the following interpretation :
as in dimension $d=1$, the dynamics renormalizes at large scales
onto a directed trap model, 
so that the length $\xi(F)$ and the inverse $1/\mu(F)$ of the anomalous exponent
represent respectively the typical length along the direction of the force, and the typical barrier $1/\mu(F)$ of an elementary renormalized trap.  
 The fact that these elementary renormalized traps are 'smaller' in linear size and in depth than in the equivalent problem in dimension $d=1$, indicates that the particle uses the transverse direction to find lower barriers. 
An important issue for the future is thus to analyse more quantitatively this
effect in order to better understand the values of the exponents $a$ and $\nu$.

Another way to understand these values could be via a strong disorder renormalization analysis : we have shown in our previous work \cite{sinai2d} that 
without external force $F=0$, the dynamics is governed by an 
'Infinite Disorder Fixed Point', so we expect that in the limit of small external force $F \to 0$, the dynamics should be described by a 'Strong Disorder Fixed Point'
as in dimension $d=1$ \cite{review}.
It is thus interesting to make some comparison with the 
two-dimensional Random Transverse Field Ising
Model (RTFIM) which has been much studied via strong disorder renormalization \cite{motrunich,lin,karevski,ferenc} (see also the Monte-Carlo study \cite{pich} on the RTFIM
and the Monte-Carlo study \cite{vojta} on the two-dimensional disordered contact
process that have found evidence for infinite-disorder behaviors) : 
it turns out that the most recent values of the correlation length exponent 
are around $\nu_{RTFIM} \simeq 1.25$ \cite{karevski,ferenc,vojta}, which is rather
close to the value $\nu \sim 1.29$ that we have measured here.
Although the RTFIM and the Sinai model are completely equivalent in dimension $d=1$ \cite{review}, the two-dimensional RTFIM and
 the dynamical model considered in the present paper
are not directly related : in particular, in the RTFIM, the disorder is local,
and the problem remains an isotropic two-dimensional model, whereas in the dynamical model, the random potential presents long-ranged correlations, and the external force
introduces a very strong anisotropy since the model becomes essentially directed
in the direction of the force. From this point of view, the fact that similar values
for the exponent $\nu$ have been found seem to be purely a coincidence. 
On the other hand, it turns out that
the value $H=1/2$ of the Hurst exponent of the random potential 
that we have chosen to define the dynamical model,
is the analog of the exponent $\psi \simeq 0.5$
 measured in the two-dimensional RTFIM \cite{lin,ferenc,vojta}.
So a more detailed comparison between the strong disorder fixed points associated to both types of models is needed to determine whether the corresponding exponents are close 
just by coincidence or not.

\section*{Acknowledgements }

It is a pleasure to thank Ferenc Igloi for useful correspondence on the similarities
and differences with other strong disorder fixed points in dimension $d=2$
(see the final discussion in the conclusion).

\end{document}